\documentclass[reprint,
              aps,
              superscriptaddress,
              nobibnotes,
              nofootinbib,
              amssymb,
              amsmath,
              ]{revtex4-2}

\usepackage[caption=false]{subfig}

\usepackage[utf8]{inputenc}
\usepackage{booktabs}
\usepackage{graphicx}
\usepackage{units}
\usepackage{textgreek}

\newcommand{\ega}{e$^{-}$/\textgamma}

\newcommand{\detA}{\textit{Det-A}}
\newcommand{\tuma}{\textit{TUM93A}}
\newcommand{\comm}{\textit{Comm2}}
\newcommand{\cawo}{CaWO$_4$}

\begin{document}

\title{Observation of a low energy nuclear recoil peak in the neutron calibration data of the CRESST-III Experiment}

\newcommand{\mpi}{\affiliation{Max-Planck-Institut f\"ur Physik, D-80805 M\"unchen, Germany}}
\newcommand{\coimbra}{\affiliation{also at: LIBPhys-UC, Departamento de Fisica, Universidade de Coimbra, P3004 516 Coimbra, Portugal}}
\newcommand{\bratislava}{\affiliation{Comenius University, Faculty of Mathematics, Physics and Informatics, 84248 Bratislava, Slovakia}}
\newcommand{\hephy}{\affiliation{Institut f\"ur Hochenergiephysik der \"Osterreichischen Akademie der Wissenschaften, A-1050 Wien, Austria}}
\newcommand{\ati}{\affiliation{Atominstitut, Technische Universit\"at Wien, A-1020 Wien, Austria}}
\newcommand{\tum}{\affiliation{Physics Department, TUM School of Natural Sciences, Technical University of Munich, D-85747 Garching, Germany}}
\newcommand{\tuebingen}{\affiliation{Eberhard-Karls-Universit\"at T\"ubingen, D-72076 T\"ubingen, Germany}} 
\newcommand{\oxford}{\affiliation{Department of Physics, University of Oxford, Oxford OX1 3RH, United Kingdom}}
\newcommand{\wmi}{\affiliation{also at: Walther-Mei\ss ner-Institut f\"ur Tieftemperaturforschung, D-85748 Garching, Germany}}
\newcommand{\lngs}{\affiliation{INFN, Laboratori Nazionali del Gran Sasso, I-67100 Assergi, Italy}}
\newcommand{\gssi}{\affiliation{also at: GSSI-Gran Sasso Science Institute, I-67100 L'Aquila, Italy}}
\newcommand{\cassino}{\affiliation{also at: Dipartimento di Ingegneria Civile e Meccanica, Universitá degli Studi di Cassino e del Lazio Meridionale, I-03043 Cassino, Italy}}

\mpi
\hephy
\ati
\lngs
\bratislava
\tum
\tuebingen
\oxford
\coimbra
\wmi
\gssi
\cassino

\author{G.~Angloher}
    \mpi
\author{S.~Banik}
    \hephy
    \ati
    
\author{G.~Benato}
    \lngs
\author{A.~Bento}
    \mpi
    \coimbra
\author{A.~Bertolini}
    \mpi

\author{R.~Breier}
    \bratislava

\author{C.~Bucci} 
    \lngs

\author{J.~Burkhart}
    \hephy

\author{L.~Canonica}
    \mpi

\author{A.~D'Addabbo}
    \lngs

\author{S.~Di~Lorenzo}
    \mpi
  
\author{L.~Einfalt}
    \hephy
    \ati
    
\author{A.~Erb}
    \tum
    \wmi

\author{F.~v.~Feilitzsch}
    \tum
    
\author{S.~Fichtinger}
    \hephy
    
\author{D.~Fuchs}
    \mpi
    
\author{A.~Fuss}
    \email[Corresponding author: ]{alexander.fuss@oeaw.ac.at}
    \hephy
    \ati
    
\author{A.~Garai}
    \mpi
    
\author{V.M.~Ghete}
    \hephy
    
\author{S.~Gerster}
    \tuebingen

\author{P.~Gorla}
    \lngs

\author{P.V.~Guillaumon}
    \lngs
    
\author{S.~Gupta}
    \hephy
    
\author{D.~Hauff}
    \mpi
    
\author{M.~Ješkovsk\'y}
    \bratislava

\author{J.~Jochum}
    \tuebingen
    
\author{M.~Kaznacheeva}
    \tum

\author{A.~Kinast}
    \tum
    
\author{H.~Kluck}
    \hephy
    
\author{H.~Kraus}
    \oxford

\author{A.~Langenk\"amper}
    \mpi
    
\author{M.~Mancuso}
    \mpi
    
\author{L.~Marini}
    \lngs
    \gssi
    
\author{L.~Meyer}
    \tuebingen

\author{V.~Mokina}
    \hephy
    
\author{A.~Nilima}
    \mpi
    
\author{M.~Olmi}
    \lngs
    
\author{T.~Ortmann}
    \tum

\author{C.~Pagliarone}
    \lngs
    \cassino
    
\author{L.~Pattavina}
    \lngs
    \tum

\author{F.~Petricca}
    \mpi
    
\author{W.~Potzel}
    \tum
    
\author{P.~Povinec}
    \bratislava
    
\author{F.~Pr\"obst}
    \mpi
    
\author{F.~Pucci}
    \mpi
    
\author{F.~Reindl}
    \hephy
    \ati

\author{J.~Rothe}
    \tum
    
\author{K.~Sch\"affner}
    \mpi
    
\author{J.~Schieck}
    \hephy
    \ati

\author{D.~Schmiedmayer}
    \email[Corresponding author: ]{daniel.schmiedmayer@oeaw.ac.at}
    \hephy
    \ati
    
\author{S.~Sch\"onert}
    \tum
    
\author{C.~Schwertner}
    \hephy
    \ati

\author{M.~Stahlberg}
    \mpi
    
\author{L.~Stodolsky}
    \mpi
    
\author{C.~Strandhagen}
    \email[Corresponding author: ]{christian.strandhagen@uni-tuebingen.de}
    \tuebingen
    
\author{R.~Strauss}
    \tum
    
\author{I.~Usherov}
    \tuebingen
    
\author{F.~Wagner}
    \hephy

\author{M.~Willers}
    \tum
    
\author{V.~Zema}
    \mpi
    
\collaboration{CRESST Collaboration}
\noaffiliation

\date{March 14, 2023}

\begin{abstract}
    New-generation direct searches for low mass dark matter feature detection thresholds at energies well below \unit[100]{eV}, much lower than the energies of commonly used X-ray calibration sources. This requires new calibration sources with sub-keV energies. When searching for nuclear recoil signals, the calibration source should ideally cause mono-energetic nuclear recoils in the relevant energy range. Recently, a new calibration method based on the radiative neutron capture on $^{182}$W with subsequent de-excitation via single \textgamma-emission leading to a nuclear recoil peak at \unit[112]{eV} was proposed. The CRESST-III dark matter search operated several \cawo-based detector modules with detection thresholds below \unit[100]{eV} in the past years. We report the observation of a peak around the expected energy of \unit[112]{eV} in the data of three different detector modules recorded while irradiated with neutrons from different AmBe calibration sources. We compare the properties of the observed peaks with Geant-4 simulations and assess the prospects of using this for the energy calibration of CRESST-III detectors.     
\end{abstract}

\maketitle

\section{Introduction}
Over the past years, many experiments searching for low-mass dark matter (DM) or Coherent Elastic Neutrino-Nucleus Scattering (CE\textnu NS) achieved detection thresholds for nuclear recoils below \unit[100]{eV} \cite{2017_Nucleus, CRESST-III_2019_firstResults, 2019_EDELWEISS, 2022_EDELWEISS, 2021_CPD, 2018_CDMS}. Conventionally, the full photoabsorption of X-rays with known energy (e.g. from $^{55}$Fe decay) leading to a peak in the observed energy spectrum at a few keV is used for energy calibration. However, these peaks may be still at too high energies to provide a reliable calibration at sub-keV energies.

Furthermore, energy scales might be shifted between electromagnetic interactions and nuclear recoils, e.g.~because of energy lost to lattice defects created by nuclear recoils  \cite{Kadribasic:2020, Matti2021, Matti2022}. An optimal calibration method would therefore provide peaks in the energy deposition spectrum originating from nuclear recoils, just like the interactions we are looking for. 

The CRAB (Calibrated nuclear Recoils for Accurate Bolometry) collaboration \cite{Thulliez_2021} proposes to use the nuclear recoil caused by radiative capture of thermal neutrons as a novel way to calibrate detectors in the \unit[100]{eV} range. Especially for the tungsten isotopes present in CaWO$_4$ crystals, they predict that the calibration signals will be visible above background in a dedicated setup, exposed to the moderated neutron flux of a research reactor. Recently, the NUCLEUS and CRAB collaborations reported on the observation of such a predicted peak at \unit[112]{eV} in a dedicated measurement of a \unit[0.75]{g} prototype detector made from CaWO$_{4}$ using a specially moderated $^{252}$Cf neutron source \cite{CRAB:2022rcm}. In the following we show that this predicted signal is also observed with high significance in data recorded with three different CaWO$_4$-based detectors in the CRESST-III setup while they were exposed to neutrons from AmBe neutron sources used for calibration.

First, we summarize the relevant findings of \cite{Thulliez_2021} in section~\ref{sec:neutron_capture} and briefly describe the experimental setup (section~\ref{sec:setup}). In section~\ref{sec:simulation} we present the results of a Monte Carlo simulation of the expected signal in the CRESST setup. We then outline the acquisition and analysis of the data in section~\ref{sec:analysis}. The results are discussed in section~\ref{sec:results} and we end with a conclusion in section~\ref{sec:conclusion}.

\renewcommand{\arraystretch}{1.2}
\begin{table} [t!]
	\centering
	\caption{Properties of the potential calibration signals and the associated thermal neutron capturing on tungsten isotopes with subsequent single \textgamma-emission: Q-value $Q$, recoil energy $E_{\textrm{R}}$, natural abundance $Y$, thermal neutron capture cross section $\sigma_{n,\gamma}$, branching ratio for single \textgamma-emission $BR_{1\gamma}$, and figure of merit FOM.}
	\label{tab:nGamma_Qvalues}
		\begin{tabular}{ r  r  r  r  r } 
			\toprule
			 & $^{182}$W & $^{183}$W & $^{184}$W & $^{186}$W \\
			\midrule
			$Q$ (keV) \cite{FIRESTONE201479}\footnotemark[1] & 6190.7 & 7411.2 & 5794.1 & 5467.0 \\
			$E_{\textrm{R}}$ (eV) & 112.4 & 160.2 & 96.1 & 85.8 \\
			$Y$ (\%) \cite{FIRESTONE201479}\footnotemark[1] & 26.50 & 14.31 & 30.64 & 28.42 \\
			$\sigma_{n,\gamma}$ (barn) \cite{ENDF8_paper} & 20.31 & 9.87 & 1.63 & 37.88 \\
			$BR_{1\gamma}$ (\%) \cite{FIRESTONE201479}\footnotemark[1] & 13.936 & 5.829 & 1.477 & 0.263 \\
			FOM & 7500.6 & 823.3 & 73.8 & 283.1 \\
			\bottomrule
		\end{tabular}
\footnotetext{\url{https://www-nds.iaea.org/pgaa/egaf.html}}
\end{table}

\section{Radiative Neutron Capture on Tungsten Isotopes}
\label{sec:neutron_capture}

Due to conservation of momentum, the radiative capture of neutrons and subsequent emission of \textgamma-rays leads to a transfer of kinetic energy to the nucleus, giving the same nuclear recoil signature as for elastic DM-nucleus scattering or CE\textnu NS. If only a single \textgamma\ is emitted in the de-excitation process, the ensuing nuclear recoil is mono-energetic. For de-excitation via multi-\textgamma\ cascades, the different combinations lead to a broader distribution of recoil energies with energies lower than the single-\textgamma\ induced recoil. The nuclear recoil component related to such \textgamma-cascades following a neutron capture has been used to study the ionization yield of silicon at low energies \cite{2022_Si_recoils}.

The Q-values for neutron capture on the naturally occurring tungsten isotopes and the respective nuclear recoil energies for single \textgamma-emission calculated via $E_{\textrm{R}} = E_{\gamma}^2 / (2 M c^2)$ are summarized in Tab.~\ref{tab:nGamma_Qvalues}. Applying the definition given in \cite{Thulliez_2021}, the relative occurrence of neutron capture-events with single \textgamma-emission as figure of merit (FOM) can be calculated by multiplying the natural abundance $Y$ of the respective tungsten isotope, the thermal neutron capture cross section $\sigma_{n,\gamma}$, and the branching ratio for single \textgamma-emission $BR_{1\gamma}$. The resulting values are listed in Tab.~\ref{tab:nGamma_Qvalues}. The peak at \unit[112.4]{eV} due to the reaction $^{182}$W(n,\textgamma)$^{183}$W is consequently expected to be by far the most prominent one. In all cases, the energy of the emitted \textgamma-ray is so high, that it is expected to escape the detector and only the nuclear recoil will be observed.

\section{Experimental Setup and Detector Modules}
\label{sec:setup}

\begin{figure}[t!]
	\centering
	\includegraphics[width=.7\linewidth]{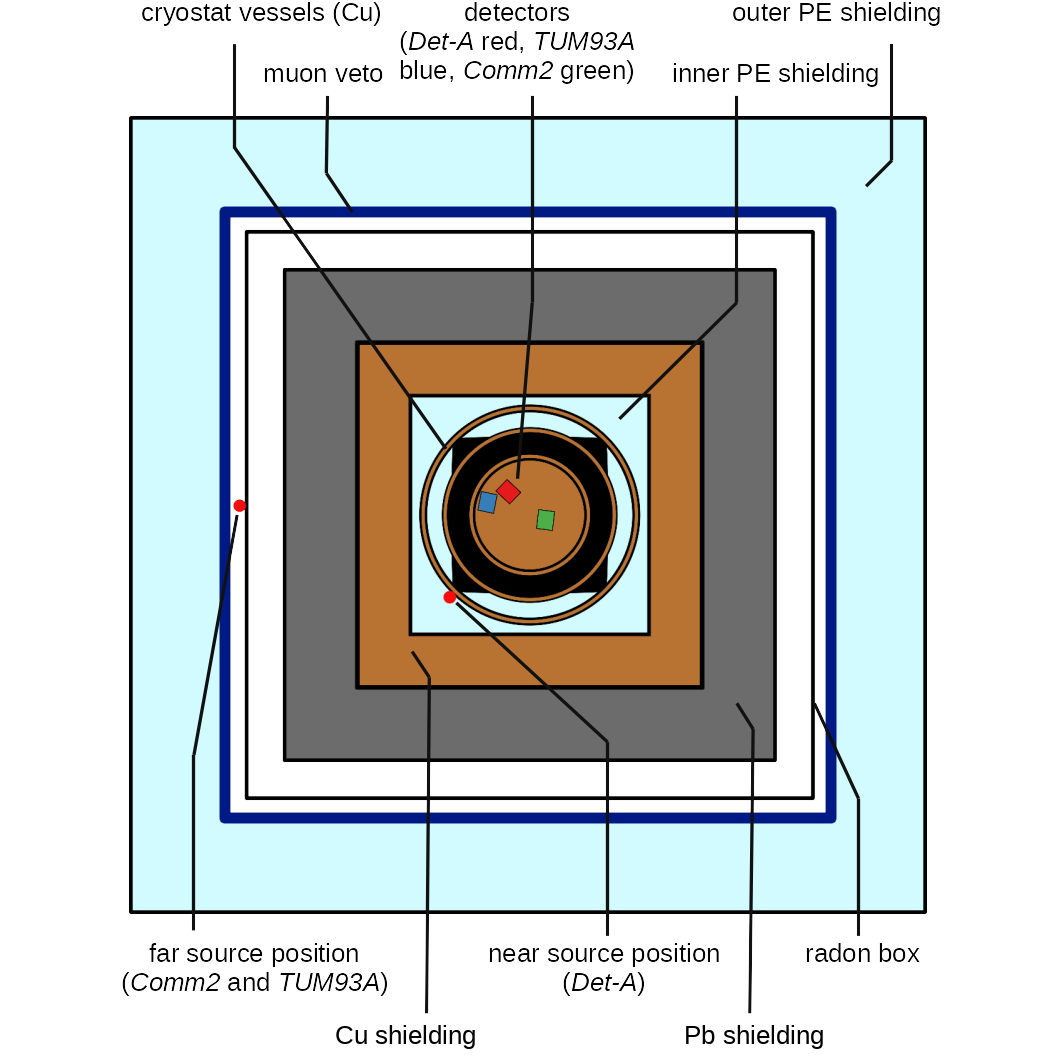}	\caption{Schematic cross-section (not to scale) of the CRESST-III setup viewed from the top. In the center, there is the copper carousel which holds the detector modules. The three modules used in this work are shown in their respective positions (red \detA, blue \tuma\ and green \comm). The carousel is surrounded by various concentric copper vessels of the cryostat. On the inside of the outer cryostat and between the vessel and the copper shielding there is a layer of polyethylene shielding (light blue). The entire setup is surrounded by box-shaped copper and lead shields and an air-tight box to keep out radon gas. The outer layers consist of an active plastic-scintillator-based muon veto and an outer layer of polyethylene. The red dots indicate the two different positions used for the neutron sources. }
	\label{fig:drawing}
\end{figure}

The CRESST experiment is located in the Laboratori Nazionali del Gran Sasso (LNGS) underground laboratory in Italy. An array of detector modules is operated in a cryostat which is surrounded by different layers of passive shielding. These are, from outside to inside, \unit[45]{cm} of polyethylene (PE), \unit[20]{cm} of lead, \unit[14]{cm} of low-background copper and additional layers of PE close to the detectors (cf.~Fig.~\ref{fig:drawing}). A more detailed description of the setup including drawings can be found in \cite{2009_Commissioning}.

The data discussed here were obtained with three different standard CRESST-III detector modules which consist of a (20\,x\,20\,x\,10)\,mm$^3$ CaWO$_4$ target crystal with a mass of \unit[$\sim$24]{g} and a (20\,x\,20\,x\,0.4)\,mm$^3$ thin silicon-on-sapphire wafer, which acts as a light detector. Both the main absorber and the light detector are equipped with a transition edge sensor (TES) operated at \unit[$\sim$15]{mK} and read out with SQUID-based electronics \cite{2007_CRESST_Squid}. On the target and light detector crystals there is an ohmic heater which is used to stabilize the sensor in its operating point in the superconducting transition. Pulses with several discrete energies are periodically injected into the heater to calibrate the response of the TES.

One of the modules - \detA\ - was operated from 2016 to 2018 and is described in \cite{CRESST-III_2019_firstResults}, the other two - \comm\ and \tuma\ - were operated from 2021 to 2022 and are described in \cite{2022_LEE_CRESST}. While the geometry is the same for all three modules, the  holding scheme is slightly different: \detA\ is held by three instrumented \cawo\ sticks, \tuma\ only has one instrumented \cawo\ stick and two copper sticks and \comm\ is held in place by bronze clamps instead of sticks. Also \comm, in contrast to the two other modules, has no reflective foil on the inside of the module housing.

\section{Neutron Calibration Simulation}
\label{sec:simulation}

To assess if a suitable fraction of the neutrons from the AmBe source thermalize within the CRESST setup, we performed Monte Carlo simulations with the ImpCRESST code \cite{CRESST:2019oqe}, which is based on Geant4 10.6.3 \cite{Allison2006,Allison2016,agostinelli2003geant4}, and incorporates, among other things, detailed geometries of the detector modules and the passive neutron shields. The datasets for this analysis come from two different measurement campaigns where different neutron sources were used in different positions (see Fig.~\ref{fig:drawing}): For \detA\ the AmBe source with a neutron output of \unit[$\sim$50]{s$^{-1}$} was put on the outside of the outer copper vessel of the cryostat, inside the copper, lead and outer PE shielding at a distance of \unit[$\sim$28]{cm} to the detector modules. For \comm\ and \tuma, which were operated at the same time, a stronger source with a neutron output of \unit[$\sim$2000]{s$^{-1}$} was placed on the radon box at a distance of \unit[$\sim$75]{cm} to the detectors. This position is outside the copper and lead shieldings, but still within the outer PE shield. To compensate for the smaller solid angle and the additional shielding material we used a stronger source compared to the near position for \detA. This resulted in similar count rates of around \unit[6]{h$^{-1}$} between \unit[0.1]{keV} and \unit[5]{keV} in all three datasets. 

\begin{figure}[t!]
	\centering
	\includegraphics[width=\linewidth]{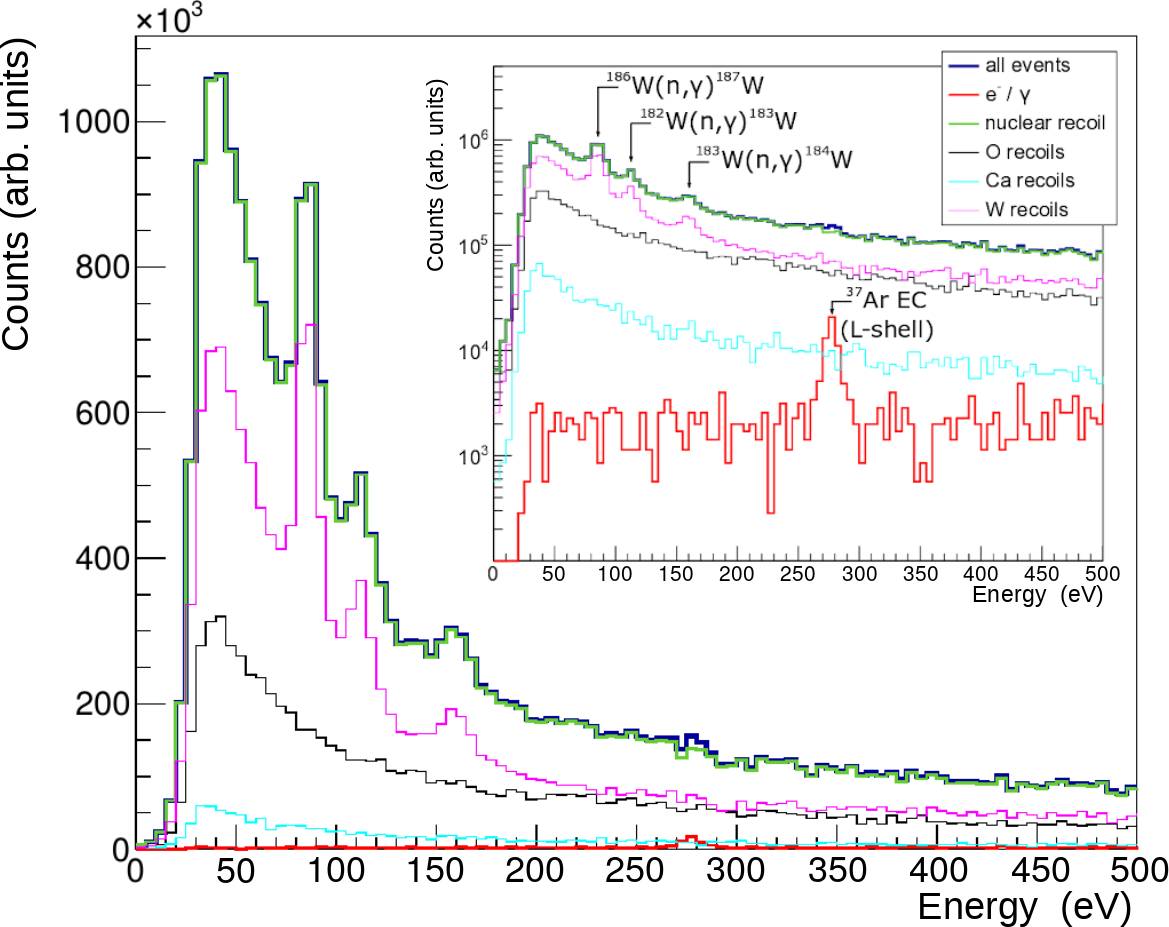}
	\caption{Geant4 simulation of energy depositions in the \cawo\ target crystal of \detA\ during a neutron calibration with an AmBe source. Energy resolution and signal survival probability as given in \cite{CRESST-III_2019_firstResults} are applied. The inset shows the same spectra with a logarithmic y-axis. The curve labeled \textit{nuclear recoil} is the sum of O, Ca and W recoils, \textit{all events} is the sum of nuclear recoils and \ega. \textit{Note:} due to limitations of Geant4, the intensities of the labeled (n,\textgamma) peaks are too high. For details see text.}
	\label{fig:nCal_Sim_Edep0to500eV}
\end{figure}

For the simulation this poses a problem, however, because significantly more neutrons have to be simulated for the far position in order to have comparable statistics to the near position. For this result, we could therefore only extract a nuclear recoil spectrum with adequate statistics for the near position for \detA\ while for the other neutron calibration campaign including \comm\ and \tuma\ we can only make a statement about the energies of the neutrons impinging on the detectors. In both arrangements, we find that a considerable amount of neutrons emitted from the source is efficiently moderated when reaching the detectors. In the simulation of the near source position, approximately \unit[21.5]{\%} (\unit[13.5]{\%}) of the neutrons impinging on the \cawo\ target crystals have energies below \unit[10]{eV} (\unit[100]{meV}). For the far position, these fractions are even higher: \unit[55]{\%} (\unit[32]{\%}) below \unit[10]{eV} (\unit[100]{meV}), because of the additional lead and copper between the neutron source and the detectors. Hence, in both cases a significant fraction of neutrons may contribute to (n,\textgamma) reactions. 

The simulated spectrum of energy depositions in \detA\ below \unit[500]{eV} shown in Fig.~\ref{fig:nCal_Sim_Edep0to500eV} is dominated by elastic nuclear recoils off tungsten or oxygen caused by fast neutrons. This component can be empirically modeled with an exponentially decaying and a constant part. On top of this, we can see the nuclear recoil peaks predicted from the radiative neutron capture processes. However, as already reported in \cite{Thulliez_2021}, we find that Geant4 treats the nuclear de-excitation after the neutron capture incorrectly: instead of considering the different possibilities of emitting multiple gammas with their respective branching ratios, in Geant4 always only a single gamma is emitted to de-excite the tungsten nucleus after a neutron capture. Therefore, the simulated spectrum can be used to predict the positions of these peaks as potential calibration signals, but not to predict their intensities\footnote{We note that in \cite{Thulliez_2021} the non-public code FIFRELIN was used to correct this shortcoming of Geant4.}. Using the branching ratios from Tab.~\ref{tab:nGamma_Qvalues} we expect that only the peak at \unit[112.4]{eV} might be observable with the data used in this work. The \ega-background induced by the neutron source, including a peak at \unit[270]{eV} from an electron capture decay of $^{37}$Ar (which is produced in (n,~\textalpha) reactions on $^{40}$Ca), is negligible. While we don't have simulated spectra for \comm\ and \tuma, we expect them to be similar to the one presented in Fig.~\ref{fig:nCal_Sim_Edep0to500eV}. Due to the higher share of thermal neutrons for the far source position, the ratio of events in the peak to those of the exponential component is likely higher for those two detectors compared to \detA.

\section{Data Acquisition and Analysis}
\label{sec:analysis}

\begin{figure}[t!]
	\centering
	\includegraphics[width=0.95\linewidth]{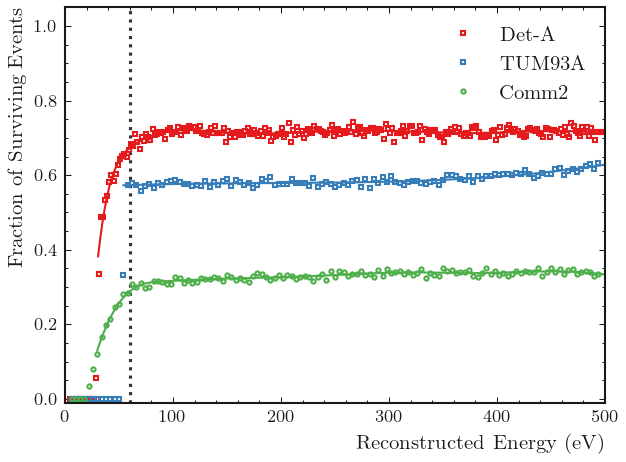}
	\caption{Fraction of simulated events surviving the trigger and event selection at a given reconstructed energy for all three detector modules (colored markers). The solid lines show the interpolation of the data points with a cubic smoothing spline, which is used to take the energy dependent survival probabilities into account. The dotted vertical line indicates the analysis threshold of \mbox{\unit[60]{eV}}. For \detA\ (red) and \comm\ (green) a dedicated cut was required to remove a special class of artifacts, which results in a decreased survival probability at low energies. For \tuma\ (blue) the survival probability is almost flat.}
	\label{fig:efficiency}
\end{figure}

\begin{figure*}[p]
    \subfloat[][\detA: light yield vs. energy plot]{
        \includegraphics[width=225pt]{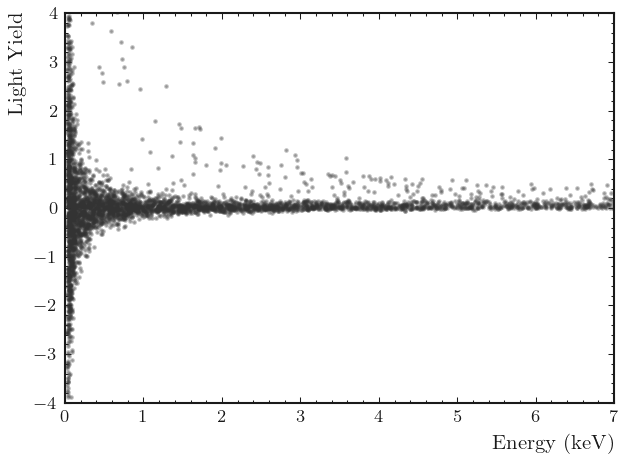}
    }
    \subfloat[][\detA: energy spctrum and fit result]{
        \includegraphics[width=0.44\textwidth]{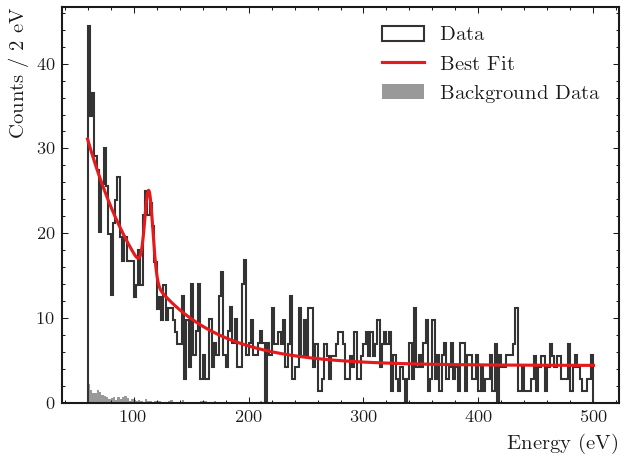}
    }

    \subfloat[][\comm: light yield vs. energy plot]{
        \includegraphics[width=0.44\textwidth]{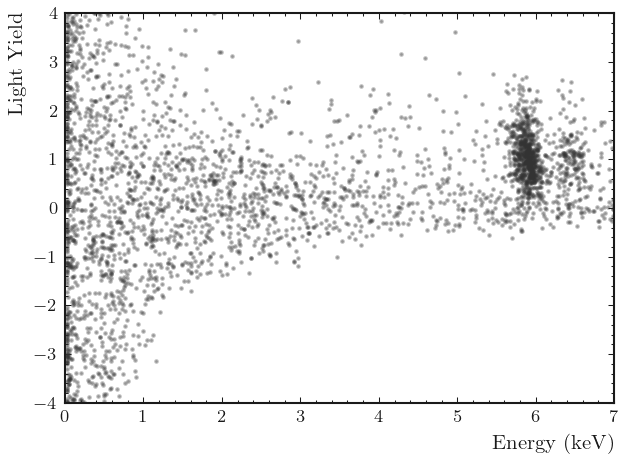}
    }
    \subfloat[][\comm: energy spectrum and fit result]{
        \includegraphics[width=0.44\textwidth]{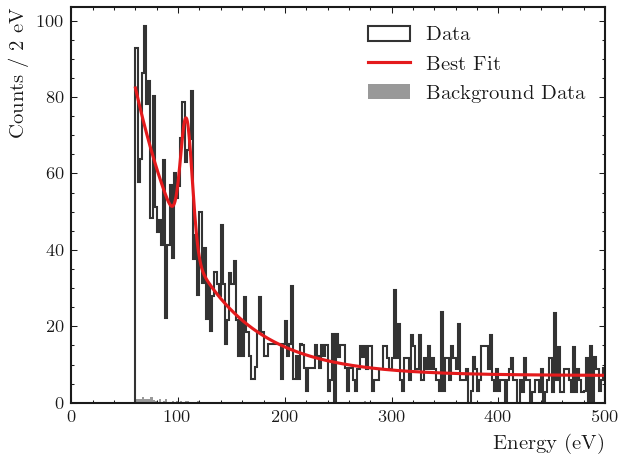}
    }
    
    \subfloat[][\tuma: light yield vs. energy plot]{
        \includegraphics[width=0.44\textwidth]{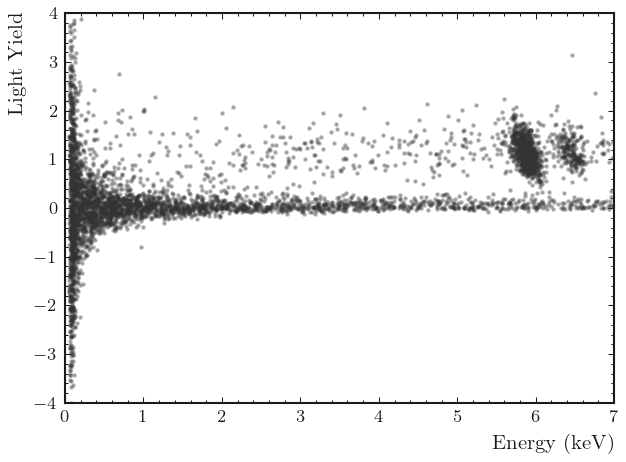}
    }
    \subfloat[][\tuma: energy spectrum and fit result]{
        \includegraphics[width=0.44\textwidth]{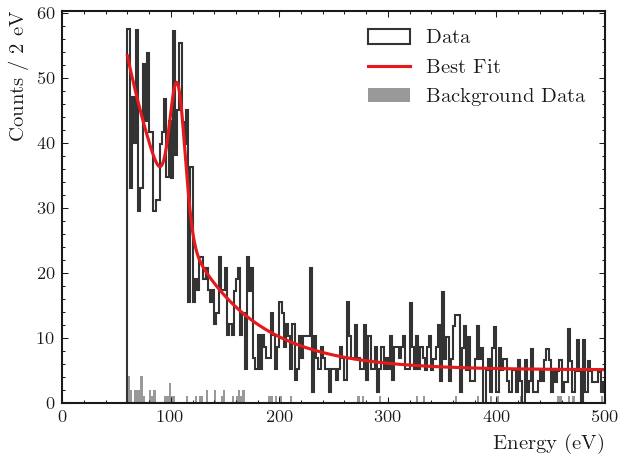}
    }
    \caption{In the left column, data from the three different detector modules are shown in the light yield vs. energy plane. For \comm\ and \tuma\ the lines from the $^{55}$Fe source and a weakly populated \ega-band are visible above the densely populated nuclear recoil band. For \detA\ there are only very few events above the nuclear recoil band. In the right column we show histograms of the energies of all accepted events between \unit[60]{eV} and \unit[500]{eV} (black line) corrected for the energy dependent survival probabilities shown in Fig.~\ref{fig:efficiency} together with the best fit curve (solid red line)} for the three different modules.We also show the energy spectra (in grey) from the background data taking without a neutron source scaled to the same exposure and corrected for the different survival probabilities. This illustrates that the count rate induced by the neutron source is much higher than the background.
    \label{fig:data-results}
\end{figure*}

A comprehensive explanation of the entire data acquisition and analysis chain for \detA\ can be found in \cite{CRESST-III_2019_firstResults}. The analysis of the other two modules follows the same procedure. In the following we sketch the most important steps.

We record the full data stream with a sampling rate of \unit[25]{kHz} and apply an offline trigger algorithm based on an optimum filter, which takes into account the signal pulse shape and the noise of the detector \cite{Gatti:1986cw}. The trigger threshold is set with the procedure defined in \cite{2018_Threshold}, such that we expect one noise trigger in an exposure of \unit[1]{kg\,d}. For small pulses in the linear regime of the TES we use the same optimum filter to determine the pulse amplitude. For larger pulses which become saturated due to the non-linear response of the TES, we use a so-called truncated standard event fit \cite{CRESST-III_2019_firstResults} for amplitude reconstruction. 

The response of the detector is linearized with the aforementioned heater pulses. To finally convert the relative energy scale to absolute deposited energy we simply use a line with a known energy. Usually these are \textgamma-lines from dedicated external calibration sources. In order to penetrate the vessels of the cryostat, sources deployed outside the cryostat need to have \textgamma-energies above \unit[$\sim$100]{keV}. For \detA\ we used a $^{57}$Co source with a prominent \textgamma-line at \unit[122]{keV} which, however, was too far above the linear range of the TES. For an approximate calibration we therefore used the tungsten K$_{\alpha}$ escape peak at \unit[63.2]{keV} and later refined the energy scale using the \unit[11.27]{keV} peak originating from an electron capture decay of $^{179}$Ta which results from cosmogenic activation of $^{182}$W. For \comm\ and \tuma\ we used weak $^{55}$Fe sources installed in the detector housing, providing X-rays with energies of \unit[5.89]{keV} and \unit[6.49]{keV}. This enables a much more reliable calibration, but comes with the downside that the source remains in the setup during the entire measurement, causing additional low energy background. In both cases, the energy used for calibration is far above the relevant energy range below \unit[1]{keV}.

For each energy deposition $E_{\textrm{p}}$ in the main absorber, we also determine the energy $L$ deposited by the scintillation light from \cawo\ in the light detector at the same time. We then calculate the light yield $LY = L/E_{\textrm{p}}$ for each event, which is set to $LY = 1$ for events from the \textgamma-line used for calibration through our calibration procedure. Due to quenching, the mean LY for nuclear recoils is lower than the one for electron- or \textgamma-events. The finite resolutions of the light and phonon detectors lead to distinct bands in the energy vs. LY plane for \ega-events, \textalpha-events and recoils from each of the nuclei in the target crystal. At low energies these bands overlap almost completely because of the small amount of scintillation light produced and the associated large relative statistical fluctuations of $L$.

We use the energy deposited in the phonon channel $E_{\textrm{p}}$ directly as a measure of the total deposited energy. This neglects a small anti-correlation between the phonon energy $E_{\textrm{p}}$ and the light yield. In the past we have shown, that this can in principle be corrected for, leading to slightly lower energies (in the order of a few percent) assigned to events with a light yield smaller than 1 (the mean light yield of calibration events) \cite{CRESST-III_2019_firstResults}. One can see the effect in the data of \tuma, where the $^{55}$Fe line around \unit[6]{keV} appears tilted in the light yield vs. energy plot (cf.~Fig.~\ref{fig:data-results}). This tilt is only visible when the light detector has a good resolution, for \comm\ and \detA\ the performance of the light detectors is worse and the tilt of the \textgamma-lines cannot be used to apply the correction. Hence we decided to treat all three detectors the same and refrain from correcting the energies. 

For the final datasets we apply a few selection criteria to remove entire time periods or individual triggered events where the energy reconstruction is not guaranteed to work properly. We first remove periods where the detector is not in its designated operating point and periods with a trigger rate that is significantly higher than the average trigger rate (usually caused by additional electronic noise). We then remove all triggered events which don't have the expected signal pulse shape, these are mostly electronic artifacts or pile-up events. For \comm\ and \tuma\ we also removed events coincident with the muon veto and/or other detector modules.\footnote{During the operation of \detA\ the muon veto could not be continuously operated.}

The probability of these selection criteria to remove valid signal events is estimated with a dedicated simulation, where we add signal templates scaled to different energies to the raw data stream and process them with the same analysis pipeline as the real data. We then determine the fraction of events being triggered and surviving all selection criteria as a function of the reconstructed energy. In Fig.~\ref{fig:efficiency} we show the resulting survival probabilities for all three detectors. For the likelihood fit we interpolate the data points with a smoothed cubic spline starting from the threshold energy of the respective detector and use this as the energy dependent function $\epsilon(E)$. The survival probability for \tuma\ is approximately constant over the considered energy range. For \detA\ and \comm\ an additional selection criterion leads to a decrease of the survival probability below \mbox{\unit[100]{eV}} towards the threshold energy.

The energy thresholds for the three modules are given together with the baseline energy resolution and the raw measurement time in Tab.~\ref{tab:eff-resolution}. As explained above, the energy threshold is set at a constant value of the optimum filter amplitude. We then simply convert this value to energy, therefore we quote no statistical uncertainties. To determine the baseline resolutions, we added a noiseless signal template scaled to a fixed energy of \unit[112]{eV} to randomly selected noise traces and apply the energy calibration procedure. We then fit a Gaussian to the distribution of the reconstructed energies, the width of this Gaussian is taken as the baseline resolution $\sigma_{\mathrm{BL}}$. This takes into account the baseline noise as well as small variations of the detector response over time and gives a lower bound of the energy resolution at low energies. 

\renewcommand{\arraystretch}{1.2}
\setlength{\tabcolsep}{7pt}
\begin{table} [t!]
	\centering
	\caption{Energy threshold $E_{\textrm{thr}}$, baseline energy resolution $\sigma_{\textrm{BL}}$ and raw measurement time $T$ for the three datasets.}
	\label{tab:eff-resolution}
		\begin{tabular}{ r  r  r  r r} 
			\toprule
			 Module & $E_{\textrm{thr}}$\,(eV) & $\sigma_{\textrm{BL}}$\,(eV) &  $T$\,(h)\\
			\midrule
			\detA & 30.6 & 4.83\,$\pm$\,0.02 & 671\\
			\comm & 29.8 & 4.95\,$\pm$\,0.02 & 862\\
			\tuma & 54.0 & 7.92\,$\pm$\,0.03 & 862\\
			\bottomrule
		\end{tabular}
\end{table}

The left column of Fig.~\ref{fig:data-results} shows the data from the neutron calibration below \unit[7]{keV} of the three different modules in the light yield vs. energy plane. In all three datasets, the nuclear recoil band can be seen as the most prominent feature. In \comm\ and \tuma\ one can also observe the lines from the $^{55}$Fe calibration source and a weakly populated \ega-band above the nuclear recoil band. \detA\ doesn't have an $^{55}$Fe source and thus has almost no visible \ega-background in the short exposure of the neutron calibration. The \ega-band in \comm\ is much wider compared to the ones in \detA\ and \tuma. This is because of the poor performance of the associated light detector. In the relevant energy range below \unit[500]{eV} the separation between the bands is not possible for all three detectors because of the low amount of light produced at these energies.

\setlength{\tabcolsep}{10pt}
	\renewcommand{\arraystretch}{1.4}
	\begin{table*} [t!]
	\centering
	\caption{Best fit parameters for the three datasets together with the statistical significance for the peak.}
	\label{tab:Results}
	\resizebox{\textwidth}{!}{%
		\begin{tabular}{rrrrrrrr} 
			\toprule
			Detector & $R_{\textrm{sig}}$\,(1/d) & $\mu$\,(eV) & $\sigma$\,(eV) & $R_{\textrm{exp}}$\,(1/d) & $E_0$\,(eV) & $R_{\textrm{flat}}$\,(1/d) & Significance \\
			\midrule
			\detA & $1.6 \pm 0.6$ & $113.3 \pm 1.4$ & $3.5 \pm 0.8$ & $25.1 \pm 2.6$ & $52.0 \pm 9.3$ & $36.8 \pm 2.4$ & $3.6\sigma$ \\
			\comm & $6.7 \pm 1.5$ & $107.8 \pm 1.3$ & $5.5 \pm 1.1$ & $58.7 \pm 3.6$ & $58.4 \pm 5.4$ & $48.0 \pm 3.1$ &  $5.8\sigma$ \\
			\tuma & $5.5 \pm 1.2$ & $106.2 \pm 1.8$ & $7.0 \pm 1.4$ & $38.0 \pm 2.4$ & $60.4 \pm 5.3$ & $34.6 \pm 2.0$ & $6.6\sigma$ \\
			\bottomrule
		\end{tabular}
	}
\end{table*}

\section{Results}
\label{sec:results}

The final energy spectra (corrected for the energy dependent survival probability shown in Fig.~\ref{fig:efficiency}) in the relevant energy range between \unit[60]{eV} and \unit[500]{eV} are shown in the right column of Fig.~\ref{fig:data-results}. In all three datasets one can see already by eye a peak at \unit[$\sim$110]{eV} on top of the exponential background. In the background data without a neutron source no peak is observed around this energy and also the background is much lower. We therefore conclude that both the peak and the exponential background must be caused by the neutron source, which also agrees with the simulations shown in Sec.~\ref{sec:simulation}. 

We determine the position, width and intensity of the peak with an extended unbinned maximum likelihood fit implemented in Python using the zfit package\footnote{\url{https://github.com/zfit/zfit}} \cite{2020_zfit}. The background is empirically modeled by the sum of an exponential with decay constant $E_0$ and a constant. For the signal we take a Gaussian with mean $\mu$ and width $\sigma$, where $N_{\textrm{x}}$ is the number of counts for the respective component:
\begin{equation}\label{eq:fitmodel}
     f(E) = N_{\textrm{sig}}\cdot \mathcal{G}(E; \mu, \sigma) + N_{\textrm{exp}}\cdot \exp(- E / E_0) + N_{\textrm{flat}}
\end{equation}

The energy dependent survival probability is taken into account in the fit by weighting each datapoint with the inverse of the interpolated curve $\epsilon(E)$ evaluated at the energy of the respective event. From the fitted count numbers $N_{x}$, we calculate the corresponding event rates per day $R_{x} = N_{x} / T$, taking into account the measurement time $T$. The resulting parameters and the statistical significance of the peak are compiled in Tab.~\ref{tab:Results}. The significance is calculated using the method described in \cite{2010_AsymptoticSignificance} as implemented in the hepstats package\footnote{\url{https://github.com/scikit-hep/hepstats}} \cite{2022_hepstats}. In all three datasets the fit finds a peak within \unit[5]{\%} of the predicted value of \unit[112.4]{eV} (at \unit[$(113.3\,\pm\,1.4)$]{eV}, \unit[$(107.8\,\pm\,1.3)$]{eV} and \unit[$(106.2\,\pm\,1.8)$]{eV}) with high statistical significance of $3.6\sigma$, $5.8\sigma$ and $6.6\sigma$ for \detA, \comm\ and \tuma, respectively. This shows that despite the relatively high energies of the calibration sources the energy calibration procedure is quite accurate down to the \unit[100]{eV} regime. The less prominent peaks expected at \unit[160]{eV} and \unit[86]{eV} can not be seen with the available statistics.

Within the current uncertainties of our energy calibration at low energies, we can not make a definitive statement if the nuclear recoil and electron recoil energy scales differ. However, the relatively small difference ($<$\,\unit[5]{\%}) of the observed and expected peak positions shows that a possible effect for tungsten recoils around \unit[100]{eV} is not very strong.

For all detectors the widths of the fitted peaks agree within uncertainties with the baseline energy resolutions. We also find statistically compatible values for $E_0$, which describes the shape of the exponential background spectrum, for all three datasets. The count rates for the nuclear recoil peak (which are corrected for the different survival probabilities) found in the datasets of \comm\ and \tuma, which are from the same neutron calibration campaign, are compatible within the statistical uncertainties. This is expected, because both detectors have the same geometry, are made of the same material and were exposed to the same neutron flux.

The ratio of the rate of events in the peak compared to the overall background is lower for \detA\ compared to the other two. This can be understood because of the different positions of the neutron source in the different calibration campaigns, which lead to different energy spectra of the neutrons reaching the detectors as outlined in Sec.~\ref{sec:simulation}. This also explains the higher significance of the peak in the datasets of \comm\ and \tuma\ compared to \detA.

\section{Conclusion}
\label{sec:conclusion}

We observe a nuclear recoil peak induced by radiative neutron capture on $^{182}$W in the data from three different \cawo\ detector modules recorded during irradiation with neutrons from AmBe sources in the CRESST-III setup at LNGS. The same peak has also recently been observed in a measurement of a prototype detector of the NUCLEUS experiment \cite{CRAB:2022rcm}. This opens up the possibility to use this feature to directly calibrate the nuclear recoil energy scale of \cawo\ detectors in the \unit[100]{eV} regime with neutrons. With an optimized neutron source it might be possible to improve the signal to background ratio and use the observed peak as the main calibration source removing the need for dedicated low energy X-ray sources in the vicinity of the detectors. While we have no indication of a noticeable position dependence from past measurements, another advantage of using neutrons is that they probe the entire volume of the absorber crystals, whereas X-ray sources only penetrate a small region close to the surface. 

The peak can also be used to study potential differences between the energy scales of nuclear recoils and electrons or gammas. With the current uncertainties of the standard energy calibration, we cannot exclude such a difference. However, if it exists it is smaller than a few percent. To improve on this, a precise calibration method using either electrons or photons at low energies needs to be developed.

\begin{acknowledgements}
  We are grateful to LNGS for their generous support of CRESST. This work has been funded by the Deu\-tsche Forschungsgemeinschaft (DFG, German Research Foundation) under Germany’s Excellence Strategy - EXC 2094–390783311 and through the Sonderforschungsbereich (Collaborative Research Center) SFB1258 “Neutrinos and Dark Matter in Astro- and Particle Physics,” by BMBF Grants No. 05A20WO1 and No. 05A20VTA and by the Austrian science fund (FWF): Grants No. I5420-N, No. W1252-N27, and No. FG1, and by the Austrian research promotion agency (FFG), Project No. ML4CPD. J. B. and H. K. were funded through the FWF Project No. P 34778-N ELOISE. The Bratislava group acknowledges a partial support provided by the Slovak Research and Development Agency (Projects No. APVV-15-0576 and No. APVV-21-0377). The computational results presented were partially obtained using the Max Planck Computing and Data Facility (MPCDF) and the CBE cluster of the Vienna BioCenter.
\end{acknowledgements}

\bibliography{main}   

\end{document}